\pdfoutput=1

\documentclass[journal,12pt,onecolumn,draftclsnofoot,]{IEEEtran}

\IEEEoverridecommandlockouts
\usepackage{changes}
\definechangesauthor[name={vahid}, color=red]{1}
\usepackage[export]{adjustbox}
\usepackage{cite}
\usepackage{amsmath,amssymb,amsfonts}
\usepackage{mathrsfs}
\usepackage{algorithmic}
\usepackage{graphicx}
\usepackage{textcomp}
\usepackage{xcolor}
\usepackage{url}
\usepackage{float}
\usepackage{amsmath}

\DeclareMathOperator*{\argmin}{arg\,min}
\usepackage{hyperref}
\usepackage{float}
\def\BibTeX{{\rm B\kern-.05em{\sc i\kern-.025em b}\kern-.08em
    T\kern-.1667em\lower.7ex\hbox{E}\kern-.125emX}}
\begin{document}

	\title{\textbf{Pilot Pattern Design for Deep Learning-Based Channel Estimation in OFDM Systems}}
	
	\author{ \IEEEauthorblockN{Mehran Soltani,  Vahid Pourahmadi, Hamid Sheikhzadeh\\
	Department of Electrical Engineering, Amirkabir Universiy of Technology, Tehran, Iran}}

	
	\maketitle

\begin{abstract}
	In this paper, we present a downlink pilot design scheme for Deep Learning (DL) based channel estimation (ChannelNet) in orthogonal frequency-division multiplexing (OFDM) systems. Specifically, in the proposed scheme, a feature selection method named Concrete Autoencoder (ConcreteAE) is used to find the most informative locations for pilot transmission. This autoencoder consists of a concrete layer as the encoder and a multilayer perceptron (MLP) as the decoder. During the training, the concrete layer selects the most informative pilot locations, and the decoder reconstructs an approximate estimation of the channel. Eventually, the ChannelNet is trained on the output of the ConcreteAE aiming to reconstruct the ideal channel response. The estimation error results show that this approach outperforms the previously presented ChannelNet with a uniformly distributed pilot pattern, and its performance is comparable to the minimum mean square error (MMSE).
\end{abstract}

\begin{IEEEkeywords}
	Channel estimation, Feature Selection, Deep Learning, Image Super-resolution, Image restoration
\end{IEEEkeywords}

\section{Introduction}
OFDM modulation, due to its bandwidth efficiency and high rate data transmitting capability, has been adopted widely in wireless communication networks such as 4G and it will be one of the key building blocks of the next generation 5G systems. One of the challenging issues in OFDM systems is to efficiently estimate the channel state information (CSI) at the receiver  to guarantee the reliable signal detection.
Pilot-based channel estimation is one of the most common methods for obtaining the CSI. Pilots are some symbols placed in specific locations of the time-frequency grid which their positions and values are known for the receiver. The receiver then estimates the channel response in all of the time-frequency grid based on the effect of the channel it observes on the neighboring pilot tones. There are some conventional methods conducted for channel estimation in OFDM systems like least square (LS) and MMSE. The LS method is a simple interpolation-based approach while the MMSE has a better performance since it uses the complete channel statistics and the noise variance; however, it suffers from higher computational complexity and needs prior knowledge of the channel statistics. 

Recently, artificial intelligence and deep learning have show an outstanding performance in various applications. Focusing on channel estimation, a multi layer perceptron (MLP) network in \cite{ye2018power} has been designed for joint channel estimation and demodulation in OFDM systems. In \cite{Deepchannel}, a denoising convolutional neural network (DnCNN) \cite{Denoising}, has been proposed for beamspace channel estimation in millimeter wave (mmWave) massive multiple-input multiple-output (MIMO) system. Our previous work named ChannelNet \cite{ChannelNet}, incorporates a combination of a super-resolution network (SRCNN) \cite{SRCNN} with a DnCNN into the pilot-based channel estimation .

The accuracy of the channel estimation depends on the locations that the pilots are transmitted within the time-frequency grid. For statistical channel estimation methods, e.g. LS and MMSE, diamond-shaped pilot pattern has been proven to be optimal \cite{diamond}; However, for DL-based ChannelNet \cite{ChannelNet}, the optimal pilot pattern is not known. With this in mind, in this paper, we present a pipeline to find and select the most informative locations in the time-frequency grid to be assigned for pilots, and afterwards, train the ChannelNet network to minimize the estimation error based on selected pilots.

In the new scheme, we implement a fully DL-based network to estimate the channel response and find the near-optimal pilot pattern using a ConcreteAE \cite{ConcreteAE} cascaded with the ChannelNet. ConcreteAE is an autoencoder-based feature selection method which consists of a concrete layer (details will come next) and an MLP as the encoder and the decoder, respectively. In this network, concrete layer acts as the feature selector (based on concrete distribution \cite{ConcreteDist}), and the decoder is the reconstruction function. To simplify the pilot design and the channel estimation pipeline, we replace the interpolation function in the first stage of the ChannelNet with the decoder of the ConcreteAE. More specifically, after pilot selection, the concrete layer is removed and the encoder network is cascaded with the conventional ChannelNet to reconstruct the channel response based on the input pilot values. 


The remainder of the paper is as follows. Reviewing background materials in Section \ref{sec:chnet},
Section \ref{sec:pilotdesign} presents our new scheme for pilot design using concrete selector layer followed by the structure of the new version of ChannelNet. Section \ref{sec:sim} compares the performance  of the proposed scheme and that of the ChannelNet with the diamond-shaped pilot pattern. Finally, Section \ref{sec:conc} concludes the paper. 

\section{Deep Learning-Based Channel Estimation} \label{sec:chnet}
\subsection{ChannelNet} \label{sec:chanelimage}

\begin{figure}
	\centering
	\includegraphics[scale=1.2]{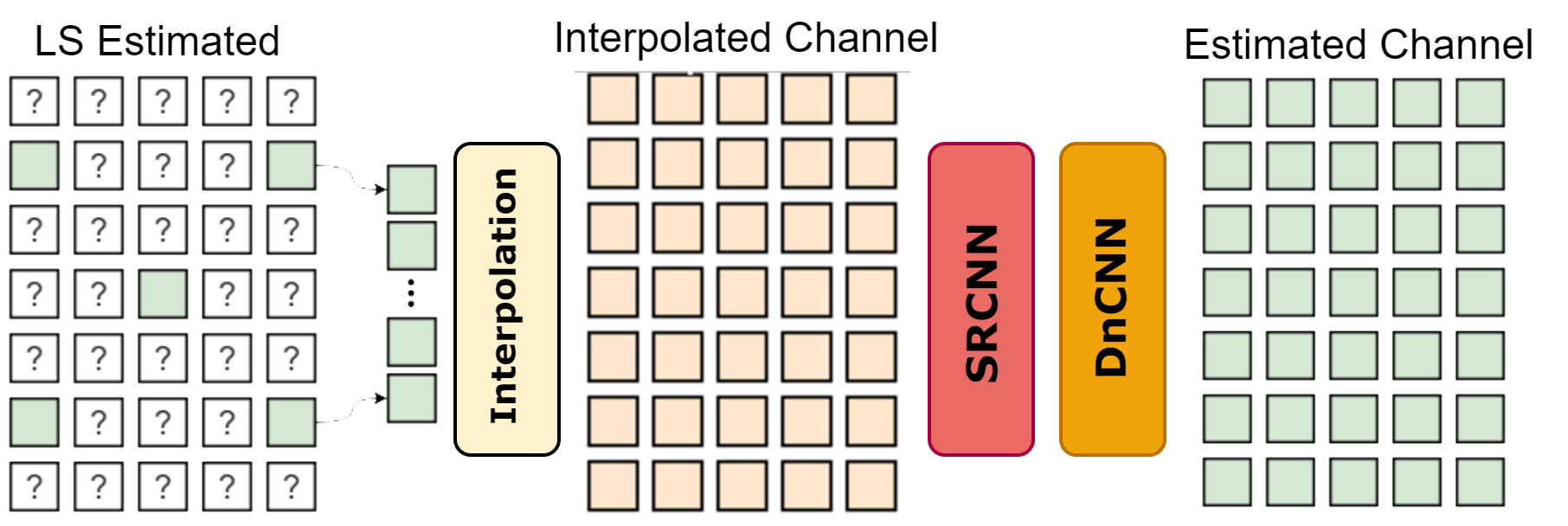}
	\caption{General pipeline of ChannelNet}
	\label{channelnet}
\end{figure}

ChannelNet is a deep neural network model proposed for channel estimation in OFDM systems (details can be found \cite{ChannelNet}). In summary, ChannelNet is a concatenation of two CNN-based networks, a super-resolution SRCNN with a denoisng DnCNN network. In this scheme, the received pilots are considered as the low-resolution image targeting at the recovery of the channel response of the whole time-frequency grid, considered as the high resolution one. Furthermore, noise effect on pilot signals is not negligible. Therefore, the DnCNN improves the performance of the SRCNN by removing the noise effect. \color{black}In ChannelNet, the input is the LS estimation at the pilot positions obtained by:

\begin{equation} \label{LS}
\mathbf{\hat{h}}_p^{\mathrm{LS}} = \argmin_{\mathbf{H}_p}  \|\mathbf{y}_p-\mathbf{\hat{h}}_p\mathbf{x}_p\|_2^2 = \mathbf{y}_p/\mathbf{x}_p,
\end{equation}
where $||.||_2$ is the $\ell_2$ distance, $\mathbf{x}_{p}$ is the transmitted signal, and  $\mathbf{y}_{p}$ is the received signal at the pilots positions. Afterwards, the LS response in  \eqref{LS} is interpolated to the size of the whole channel, and then passes through the fully-convolutional SRCNN and the DnCNN networks to estimate the whole channel response $\hat{\textbf{H}}$.
\color{black}
Considering the $f_S(.)$, $f_D(.)$  as the SRCNN and the DnCNN functions, respectively, their corresponding parameters known as $\Theta_S$ and $\Theta_D$, are optimized in a two-stage process aiming to minimize the following loss function:

\begin{equation}
C = \frac{1}{\|\cal{T}\|}  \sum_{\textbf{h}_{p} \in \cal{T}}\|\hat{\textbf{H}}-\textbf{H}\|_2^2,
\end{equation}
where $\hat{\textbf{H}}$ and  $\textbf{H}$ are the estimated and the desired channels, respectively. General pipeline of ChannelNet is shown in \figurename{\ref{channelnet}}.

\subsection{Pilot design scheme}

Although ChannelNet with equally-spaced pilots (similar to the pilot locations in LTE) demonstrated promising results, it has not been investigated whether such pilot design is the best or maybe some other pilot selections may lead to a better channel estimation. \figurename{\ref{pilotpattern}} shows two examples of equally-spaced and non-equally-spaced pilot patterns.  

For the ChannelNet, by combinatorially searching, we can obtain the optimal pilot pattern with the minimum reconstruction error. However, it is computationally prohibitive to find the optimal set of $N_p$ candidates from $N$ possible points for pilot transmission, especially when $N_p$ and $N$ are not small values. For example, we want to find location for transmission of  16 pilots, $N_p=16$ within a time-frequency grid of size $N=72\times14$, we have $ {{72\times14} \choose {16}}$ different pilot patterns, which form a huge search space. 
Therefore, in this work, we propose a practical feature selection scheme to obtain near-optimal pilot pattern for each specific channel model.

\subsection{Concrete selector layer}

ConcreteAE \cite{ConcreteAE} is a recent scheme proposed for feature selection.
Compared to the previous DL-based feature selection methods, ConcreteAE has shown a better performance in finding the most informative features and removing the most redundant ones. In this network, the selector layer consists of $k$ nodes (set by the user) and the weights specified by the Concrete random variable \cite{ConcreteDist}. This distribution is a continuous relaxation of one hot vector controlled by a temperature parameter $T \in (0, \infty)$. The selector layer generates stochastic linear combinations of the input features during the training, and smoothly converges to a discrete set of k features. To sample a $d$-dimensional Concrete random variable with the \color{black}fixed parameter $T$ and trainable parameter $\alpha \in R_{>0}^d$, \color{black}one first samples a $d$-dimensional i.i.d. vector $\textbf{m}=[m_1, m_2, ..., m_d]$ from a Gumbel distribution \cite{Gumbel}, g, and then computes each element of the Concrete distribution based on the following:

\begin{figure} 
	\centering
	\includegraphics[scale=.08]{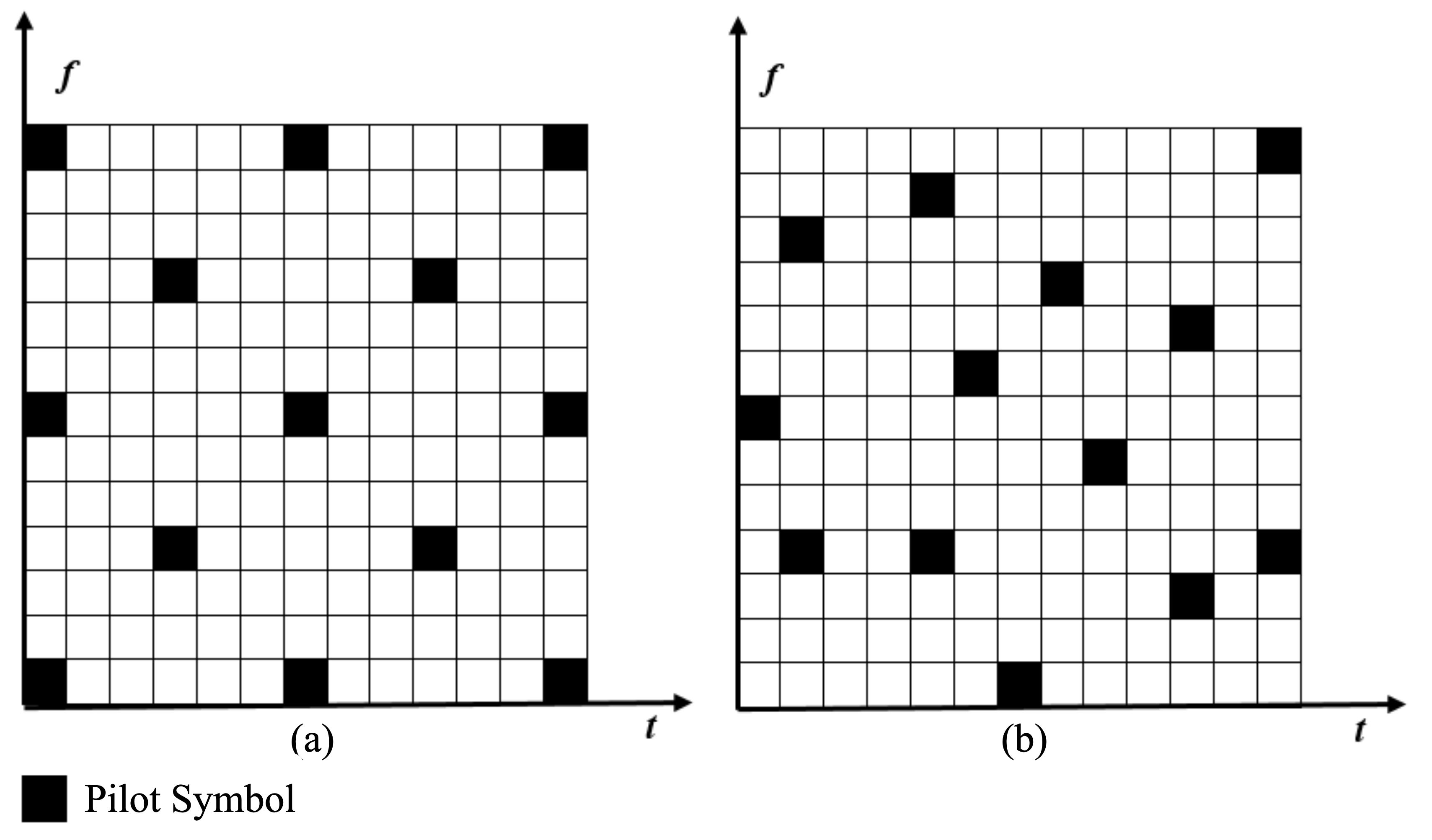}
	\caption{Examples of deterministic pilot patterns. a) equally-spaced b) non-equally-spaced}
	\label{pilotpattern}
\end{figure}

\begin{equation} \label{concrete}
m_j = \frac{exp((log\alpha_j + g_j)/T)}{(\Sigma_{k=1}^{d} exp((log\alpha_k + g_k)/T)}
\end{equation}
where $m_j$ refers to the $j$th element in the sample vector $\textbf{m}$. When $T\rightarrow 0$ \color{black}manually\color{black}, the concrete variable approaches the one hot vector $m_j$ with probability $\alpha_j / \Sigma_p \alpha_p$.

\section{Pilot Design - ChannelNet with ConcreteAE}\label{sec:pilotdesign}

According to the restricted isometry property \cite{RIP}, it has been proven that the randomly generated pilot pattern is statistically optimal for MMSE channel estimation. However, the implementation of the random pilot pattern is challenging and it is not realistic in practical scenarios due to its high complexity. Therefore, practical OFDM systems use a deterministic and equally-spaced pilot pattern.

The pilot design question is in fact investigating which of the time-frequency locations should be selected for pilot transmission such that we get the best performance of the channel estimation. In the proposed scheme, \textit{\underline{the pilot selection problem}} has been observed as \textit{\underline{a feature selection task}} and then we have used recent ideas on DL-based feature selection to find the set of pilot locations. 

	
More specifically, we have proposed a channel estimation pipeline with two specifications: \textbf{a)} A ConcreteAE is utilized to find the most informative locations for pilot transmission. \textbf{b)} Considering the selected pilots, we propose an improved version of the ChannelNet which can operate in different noise levels. In the following we describe these two parts in details.


\textbf{Pilot Pattern Design:} Similar to \cite{ChannelNet}, we consider the channel response as an image. For selecting the $N_p = k$ most informative pilot locations, \color{black} first, we vectorize the real and the imaginary part of the noisy and the ideal time-frequency channel grid and treat each part as a separate channel. Each input to the ConcreteAE is the vectorized version of the channel $\textbf{h}_{noisy} = [x_1, x_2, .. x_d]$, and the output is the reconstructed vectorized response $\hat{\textbf{h}} = [\hat{x_1}, \hat{x_2}, ..., \hat{x_d}]$, where $d$ is the length of the vectorized time-frequency grid.

\begin{figure}
			\includegraphics[width=0.8	\textwidth,center]{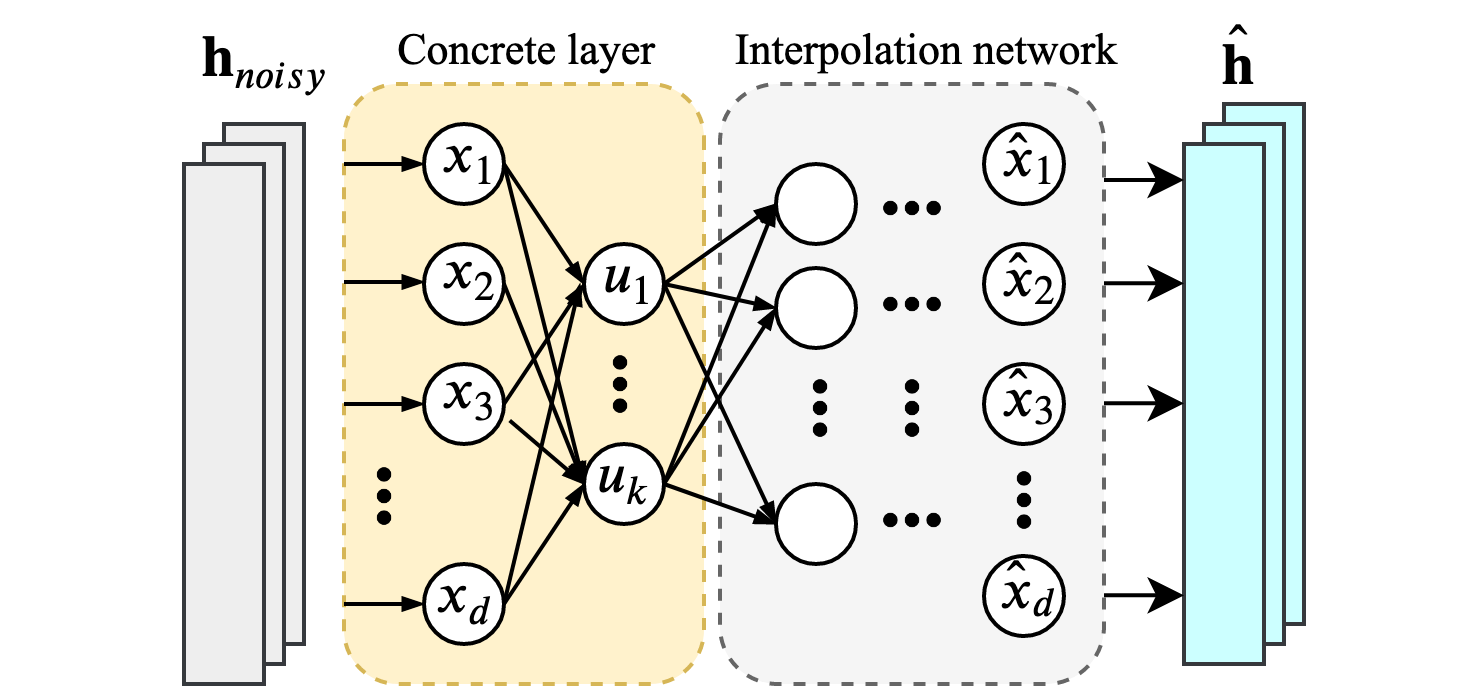}
			\caption{Training pipeline of ConcreteAE for pilot selection}
			\centering
			\label{Concrete}
		\end{figure}

For each node of the selector layer a $d$-dimensional concrete random variable $\textbf{m}_i\sim Concrete(\alpha^{i},T), i\in{1,...,k}$, is assigned. The $i$th node outputs $u_i = \textbf{h}_{noisy}.\textbf{m}_i$, and the values of $\alpha^{i}$ and the weights of the decoder are iteratively updated by minimizing the following loss function:
	
	\begin{equation}
	L = \frac{1}{N}  \sum_{n=1}^N\|f_\theta (\textbf{u}^n)-\textbf{h}_{ideal}^n\|_2 ^2,
	\end{equation}
	where $f_\theta (.)$ is the decoder function, $N$ is the number of samples, $ \textbf{u}^n\in R^{k} $ is the vector containing $u_1 , ...,u_k$ values of the  $n$th training sample $\textbf{h}_{noisy}^n$, and $\textbf{h}_{ideal}^n$ is the  $n$th corresponding vectorized ideal channel response.
	\color{black}
	 Furthermore, during the training process, the value of the temperature $T$ begins with a high value $T_0$ and gradually decreases until a final temperature $T_B$ close to zero.
	 \color{black}
	 As $T\rightarrow 0$, each node in the concrete selector layer outputs only one of the input nodes, i.e., select that feature. Figure \ref{Concrete} shows the pipeline of the proposed pilot selection approach.
		
\begin{figure}
			\includegraphics[width=1.2\textwidth,center]{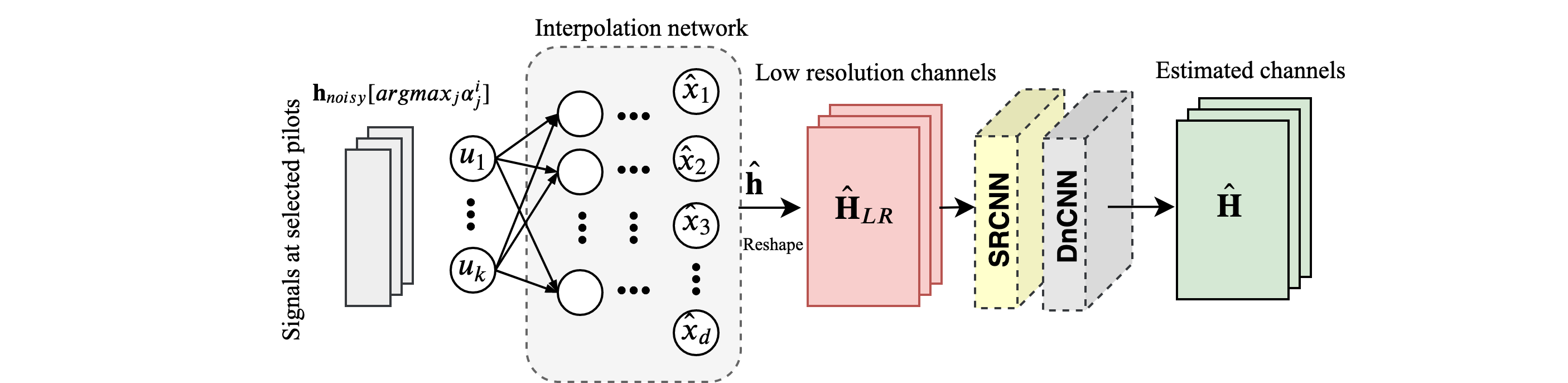}
			\caption{Training pipeline of ChannelNet based on selected pilots}
			\label{ConcChannelNet}
		\end{figure}

	\color{black}
	\textbf{Channel estimation:} After selecting the pilots,  we replace the concrete selector layer with a discrete $argmax$ layer in which the output of the $i$th neuron is  $u_i=\textbf{h}_{noisy}[argmax_j{\alpha_j^i}][i]$. So as shown in \figurename{\ref{ConcChannelNet}}, $\textbf{u}$ (signals at pilot locations), is then passed through the ConcreteAE interpolation network so we get the vectorized low resolution version of the channel estimate,  $\hat{\textbf{h}} = [\hat{x_1}, \hat{x_2}, ..., \hat{x_d}]$. Reshaping $\hat{\textbf{h}}$ to the original channel frame size  $\hat{\textbf{H}}_{LR}$, it will be fed to the ChannelNet so we get the final channel estimation $\hat{\textbf{H}}$. 
	
	There are three main differences between the new ChannelNet and the one we have proposed in \cite{ChannelNet}. First, as we discussed, the interpolation function of the original ChannelNet is replaced with the decoder of the ConcreteAE. 
	Secondly, in the new version we have used a DnCNN for blind Gaussian denoising (DnCNN-B) rather than a typical DnCNN. With DnCNN-B, ChannelNet can reconstruct the channel with a wider range of SNRs without loosing the reconstruction performance. Lastly, here, both SRCNN and DnCNN have been trained in an end-to-end scheme. This approach reduces the complexity of the proposed channel estimation network, and makes it more reliable for the practical scenarios. 
	The end-to-end loss function that we have used for training is:
	
	\begin{equation}
	L = \frac{1}{N}  \sum_{n=1}^N\|f_R(\Theta_R;f_S(\Theta_S;\hat{\textbf{H}}_{LR}^n))-\textbf{H}_{ideal}^n\|_2^2,
	\end{equation}
	
	 \section{Simulation Results}	 \label{sec:sim}	 
	 To evaluate the proposed pipeline, we consider a single antenna case and set the channel frame size to be $N_f = 72$ subcarriers and $N_n = 14$ time slots. For the channel modeling and pilot transmission, we have generated a Vehicular-A (VehA) with carrier frequency of 2.1 GHz, bandwidth of 1.6 MHz and UE (user equipment) speed of 50 km/h,
	 from LTE simulator developed by university of Vienna, Vienna LTE-A simulator \cite{Mehlfuhrer2011}. For training, testing and validation sets, 32000, 4000 and 4000 channels have been generated, respectively. To have a satisfactory performance in a wide range of SNR with a single network, we also generated the noisy version of the channels covering the SNR of 0 to 30 dB with 3 dB step size.
	 
\begin{figure}
			\includegraphics[width=0.19\textwidth,center]{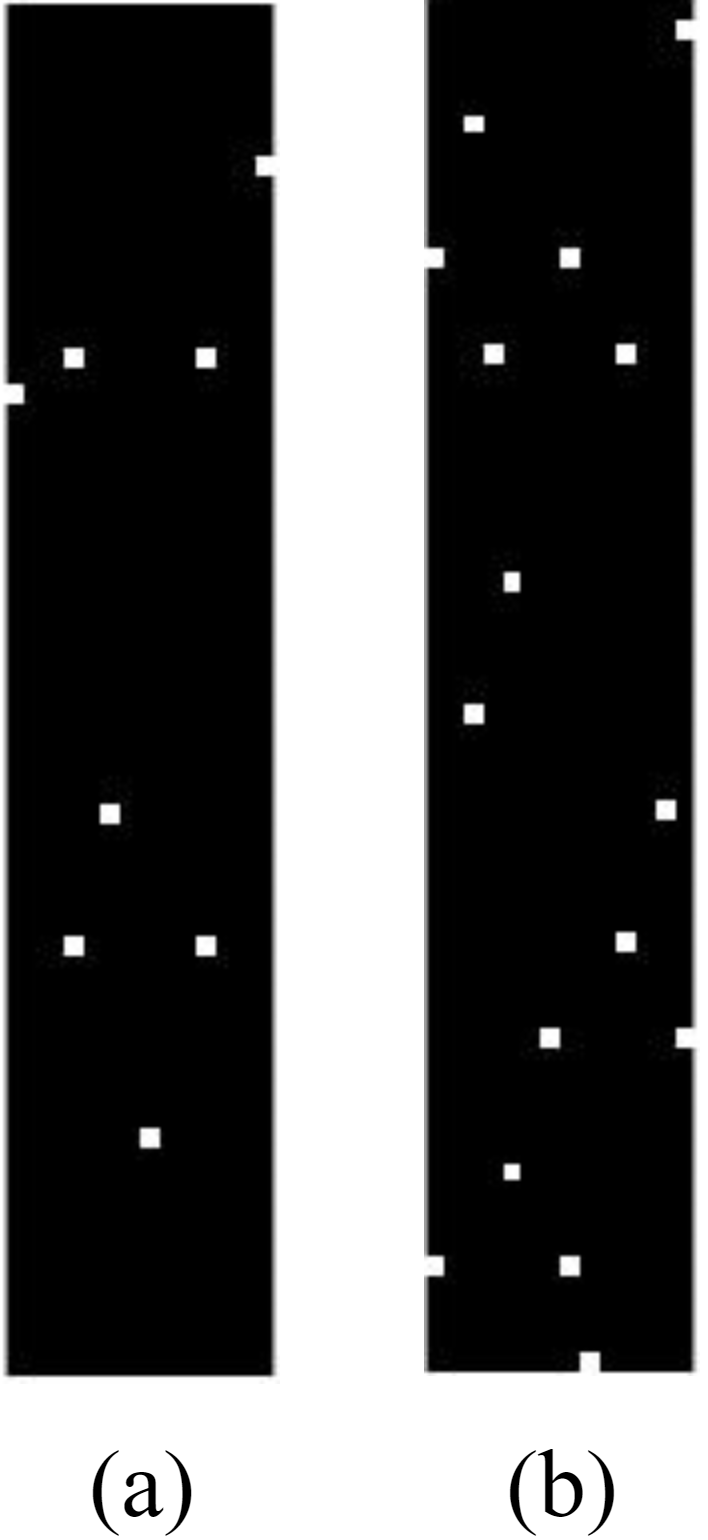}
			\caption{Designed patterns for VehA a)$N_p=8$8 and b)$N_p=16$}
			\label{patterns}
	   	\end{figure}

     For the decoder of the ConcreteAE, we have used a 3-layer MLP, each followed by a LeakyRelu(0.2) and a Dropout(0.1) (More layers did not improve the results considerably). Applying the ConcreteAE for VehA channel model, we can find the good location for pilot transmissions. \figurename{\ref{patterns}} shows the resulted pilot patterns for $N_p = 8$ and $N_p = 16$. As can be seen, pilots are almost distributed along different subcarriers rather than time slots, which seems reasonable due to the wavy nature of VehA channels in the frequency domain and almost similar values in time domain.
	
\begin{figure}[H]
				\includegraphics[width=0.72\textwidth,center]{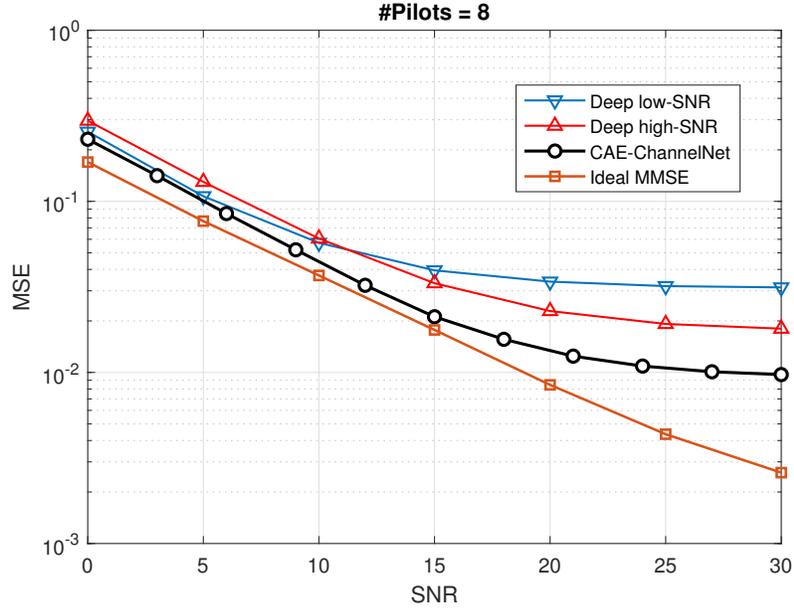}
				\caption{MSE results for 8 number of pilots}
				\label{pilot8}
			\end{figure}

\begin{figure}[H]
		 		\includegraphics[width=0.72\textwidth,center]{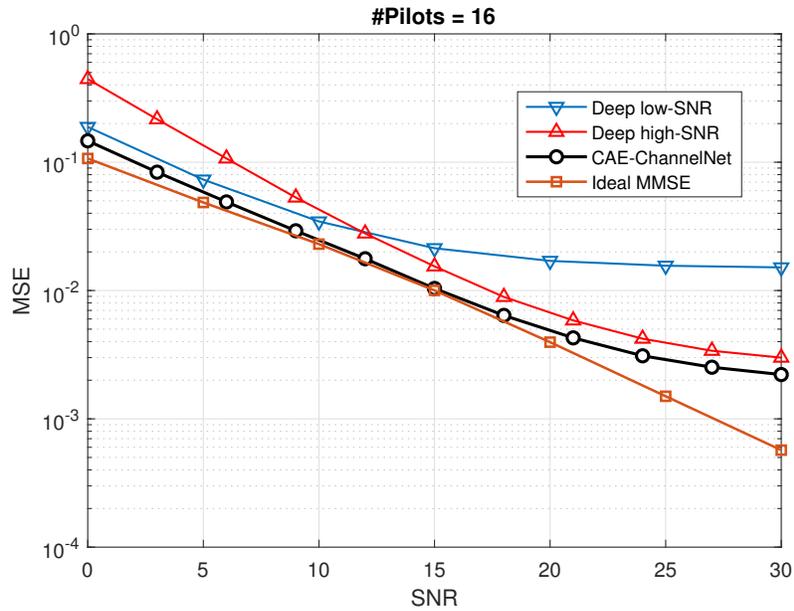}
				\caption{MSE results for 16 number of pilots}
				\label{pilot16}
			\end{figure}

	Another important evaluation is to compare channel estimation MSE of the new pilot design of ChannelNet with equally-spaced pilot. For $N_p = 8$ and $N_p = 16$, the channel estimation MSE of typical ChannelNet with uniform pilot pattern (Deep low-SNR and Deep high-SNR), ChannelNet with ConcreteAE for pilot selection (CAE-ChannelNet), and Ideal MMSE, are depicted in \figurename{\ref{pilot8}} and  \figurename{\ref{pilot16}}, respectively. The CAE-ChannelNet outperforms both Deep low-SNR and Deep high-SNR networks, especially in high SNRs, and it is comparable to the Ideal the MMSE. We remind that the ideal MMSE requires second order channel statistics and noise variance as prior information, which is impractical in real communications, therefore its performance is a lower bound. 	  

\begin{figure}[H]
	 	\includegraphics[width=0.72\textwidth,center]{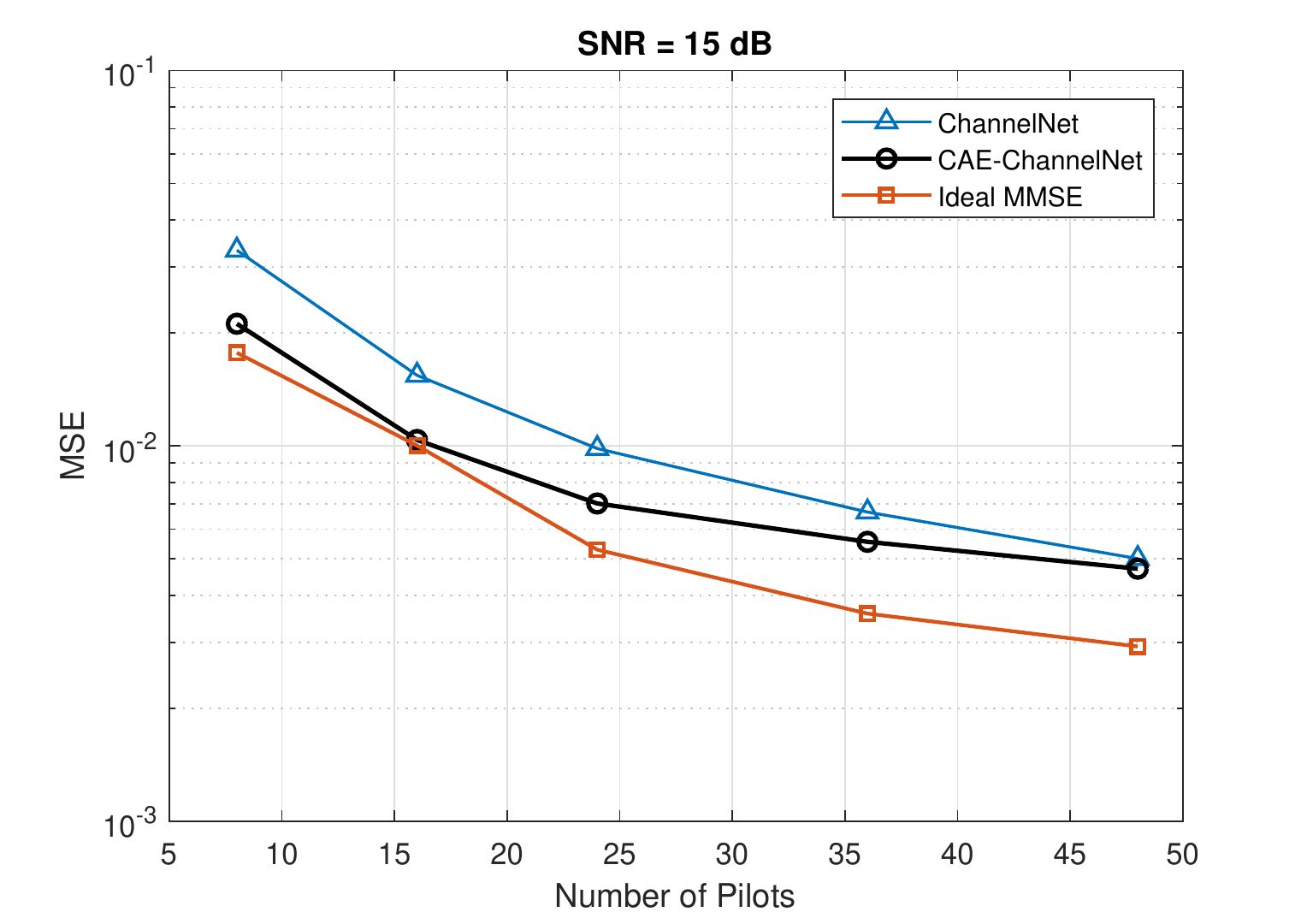}
	 	\centering
	 	\caption{MSE results based on number of pilots}
	 	\label{MSEPilots}
	 \end{figure}

			 As the final analysis, we have compared the performance of the new pilot pattern and the equally-spaced design when we change the number of available pilots. Figure \ref{MSEPilots} presents the MSE result for the VehA channel model based on $N_p$ and at the SNR value of 15dB. As can be seen, when the $N_p$ increases, MSE of CAE-ChannelNet converges to the MSE of the typical ChannelNet with the uniform pilot pattern. Therefore, for this specific channel model, using a pilot design scheme is more essential when the aim is to send a low number of pilots. For a relatively higher pilot numbers, as we have measurements from more pilot locations, even equally-spaced pattern results in an acceptable MSE. 

 \section{Conclusion} \label{sec:conc}
	In this paper we have derived a non-uniform pilot design scheme for deep learning based channel estimation in OFDM systems. First, the proposed scheme uses a concrete autoencoder to find the most informative pilots in the time-frequency grid, and afterwards, ChannelNet has been trained based on the selected pilots. Simulation results demonstrate that the proposed scheme outperforms the previously developed ChannelNet on the equally-spaced pilot pattern with only one trained network on a wide range of SNRs. 

\bibliography{ref}
\bibliographystyle{ieeetr}

\end{document}